# The CALBC RDF Triple store:
# retrieval over large literature content

Samuel Croset[1], Christoph Grabmüller[1], Chen Li[1],
Silvestras Kavaliauskas[1], Dietrich Rebholz-Schuhmann[1]

[1] EMBL Outstation, European Bioinformatics Institute,
Hinxton, Cambridge, CB10 1SD, U.K.
{Croset, Grabmuel, Chenli, Kavalia, Rebholz}@ebi.ac.uk

**Abstract.** Integration of the scientific literature into a biomedical research infrastructure requires the processing of the literature, identification of the contained named entities (NEs) and concepts, and to represent the content in a standardised way.
The CALBC project partners (PPs) have produced a large-scale annotated biomedical corpus with four different semantic groups through the harmonisation of annotations from automatic text mining solutions (Silver Standard Corpus, SSC). The four semantic groups were chemical entities and drugs (CHED), genes and proteins (PRGE), diseases and disorders (DISO) and species (SPE). The content of the SSC has been fully integrated into RDF Triple Store (4,568,678 triples) and has been aligned with content from the GeneAtlas (182,840 triples), UniProtKb (12,552,239 triples for human) and the lexical resource LexEBI (BioLexicon). RDF Triple Store enables querying the scientific literature and bioinformatics resources at the same time for evidence of genetic causes, such as drug targets and disease involvement.

**Keywords:** Triple Store, Text Mining, Data Integration

## 1  Introduction

The scientific literature is the primary data resource reporting on novel findings from the scientist. In the biomedical domain, the scientific literature is increasingly embedded into the realm of the scientific databases.  This leads to the need to interlink the content from the literature with the content from the scientific databases and to exploit both resources through the same means of access, for example using a single application for browsing all data resources or performing analyses across the data resources for consistency analyses and hypothesis testing [1, 2].

   Ideally both resources represent the same type of information.  Unfortunately, the biomedical scientific data resources represent content and its semantics that is defined by the database provider who follows either his own demands or the demands of a smaller part of the scientific community [3].  In the same way, the scientific literature conveys the semantics of the author which is not necessarily aligned with the semantics of the primary data resources [4]. As a result, the curators to the biomedical scientific data resources receive the role as gatekeepers between the literature and the primary biomedical data resources.

In recent years, significant scientific efforts have been spent to export database semantics to data representations that follow open standards and explicitly state the semantics of the content. One of the most open standards is the representation of data in RDF and the delivery of data in triple stores to achieve semantic interoperability in the Semantic Web [5,6,7,8]. This approach leads to three main advantages: First, the use of concepts and relations that are specified based on definitions available from open access resources leads to consistent reuse of content across distributed resources [9, 10]. Second, the standardised and transparent data representation improves reuse of data and error handling [5]. Third, the simplicity and generality of the data representation supports large-scale and seamless exploitation of the data [8]. Overall, the use of data across data resources requires open standards, but the scientific literature is not necessarily part of the data integration and data distribution activities.

The literature reporting on biomedical research contains scientific facts that are often subsequently integrated into biomedical data resources with the help of manual curation. This process is time-consuming and error-prone. Automatic processes for data integration of facts from the scientific literature would reduce curation efforts and would render the data transfer into a formalized process that would undergo continuous quality improvement [1]. Automatic processing of the scientific literature requires standardisation of the processes, the means and the outputs [11, 12]. A number of initiatives have been proposed to provide quality assurance to the transformation of text into database content. On the one side, annotated corpora have been made available to test text mining solutions (JNLPBA, PenBioIE, BioCreative) on the other side the curation teams work towards shared resources to standardise the outcome of their work [13,14,15,16,17]. In particular, the generation and maintenance of ontological resources form a crucial step in the development of shared semantic resources for interoperability of knowledge repositories [18,19,20].

In this manuscript we describe the combination of different data resources that have been brought together to demonstrate the benefits of semantic interoperability in the biomedical domain. The use of standardised annotations in the scientific literature, i.e. the annotations in the CALBC corpus, in combination with a lexical resource, i.e. the BioLexicon, and the data integration of both resources with publicly available data repositories, i.e. UniProt and ArrayExpress, lead the way to the exploitation of the scientific literature in the Semantic Web [21,22,23,].

## 2 Method

### 2.1 BioLexicon / LexEBI

The BioLexicon is a terminological resource that that contains references to terms from different primary data resources: BioThesaurus 6.0 (including UniProt amongst other resources), ChEBI (release 64), NCBI taxonomy, disease terms from UMLS (release 2010AA) and other data resources [24,22,25,26,27]. The terminology is kept in a standardised format in a MySQL database for the BioLexicon and in an XML formatted data repository for LexEBI [20,19]. The lexical resource serves as a complete term repository for the biomedical domain and enables disambiguation of entity types. In the case of protein and gene names (PGNs) we can disambiguate PGN terms that are polysemous with the following additional meanings: (1) the term has a meaning in general English, e.g. CAT, (2) the term serves as a hypernym, e.g. in the case of generic enzyme names (e.g. oxidoreductase), (3) the term is used for orthologous and homologous genes, and (4) the term is used with an alternative biomedical meaning, e.g. for retinoblastoma [20].

The BioLexicon contains a number of features, such as frequency counts for the occurrence of the term in the British National Corpus (BNC) or in Medline, for the number of concept ids that the term belongs to, for the number of MESH nodes that the term matches and the number of taxonomic ids that are linked to the term. All information can be used to disambiguate terms in the literature against the distribution of the term in other resources. For example, if a term is frequent in the BNC, then it tends to be less specific than another term that appears at a higher frequency across Medline. The BioLexicon contains the following number of entries (see table 1).

**Table 1.** The table gives an overview on the content from the terminological resource BioLexicon / LexEBI. Not all contained entity types are listed, i.e. the table shows only those entity types that have been integrated into the CALBC triple store. A concept id or cluster id has always a reference to the primary data resource such as UniProt or the NCBI taxonomy. The concept id makes reference to the preferred tem and the term variants. The overall number of unique terms is lower than the number of term variants due to term ambiguity across the lexical repository.

|  | # Clusters | # Variants | # Unique terms |
|---|---|---|---|
| Genes/proteins (6.0) | 488,577 | 3,389,316 | 1,564,436 |
| Chemicals (ChEBI) | 19,645 | 94,748 | 101,307 |
| Species | 643,280 | 199,130 | 838,135 |
| Diseases | 27,157 | 165,581 | 161,875 |
| Total | 1,178,659 | 3,848,775 | 2,665,753 |

## 2.2 The CALBC corpus

All project partners (PPs) of the CALBC project a corpus of 150,000 Medline abstracts with their text mining and annotation solutions. All annotations were delivered in the IeXML format and concept normalisation made use of standard resources such as UMLS, UniProtKb, EntrezGene or at least had to follow the UMLS semantic type system [11,30,25,22,26].

The alignment of the annotations for the generation of the Silver Standard Corpus I (SSC-I) is based on the methods described in [21,23]. The applied method used pair-wise alignment between two annotated sets for a given semantic type. For every sentence the annotations from one contribution for a given type is aligned with the annotations from the next contribution for the same semantic type. Different schemes for the similarity measurements have been applied to achieve the alignment, the measurements and the harmonisation of the corpus. The SSC-I has been made publicly available to enable challenge participants (CPs) to compare their annotation solutions against the annotated corpus.

**Table 2.** The table shows the number of annotations that are contained in the SSC-I [29, 31]. This corpus has been generated from the contributions of the PPs. Not all challenge participants (CPs) have participated in all parts of the challenge. A smaller number of CPs have submitted annotations for chemical entities. The average number of annotations for CHED and PRGE in the submitted corpora was above the number of annotations in the SSC-I and for DISO and SPE below the number of the ones in the SSC-I.

|      | Nr. Of anntoations in the SSC-I | Nr. Of CPs | Nr. Of submissions from CPs | Average nr. Of annotations from all CPs | Nr. Of annotations in the SSC-II |
|------|---|---|---|---|---|
| CHED | 228,622 | 6 | 11 | 233,398 | 238,431 |
| PRGE | 275,235 | 9 | 15 | 343,681 | 435,797 |
| DISO | 300,637 | 8 | 11 | 255,599 | 245,524 |
| SPE  | 317,211 | 7 | 9  | 277,071 | 304,503 |

All submissions from all CPs have been evaluated against the SSC-I and the contributions with the best F-measure performance from each CPs have been selected for the harmonisation into the SSC-II. A varying number of contributions from the CPs were available for the harmonisation of CHED, SPE, DISO and PRGE (see table 2).

The alignments of the 100,000 documents were either performed on Sun Fire opteron servers (4 or 8 CPUs, RAM sizes from 32 to 256 Gb RAM, 9-12 hours) or on the compute farm of 700 IBM compute engines (dual CPU, 1.2-2.8 Ghz, 2 GB RAM, 3 hours).

### 1.2 UniProtKb and ArrayExpress

The integration of content from GeneAtlas has not been finalized yet. GeneAtlas offers a Java API for data export. Different serialisations in XML and JSON currently export 138 experiments. The annotations of the experiments are based on the Experimental Factor Ontology (EFO) which contains concepts from a wide range of conceptual resources. All triple stores have been implemented based on Jena TDB[1].

---

[1] http://jena.hpl.hp.com/wiki/SDB

## 3 Results

### 3.1 Integration of the literature content into the triple store

The content of the scientific literature is processed to identify entities, concepts and facts from the literature (see figure 1). The content from the scientific database has to be processed to support the needs of the information extraction infrastructure for the identification and normalisation of named entities. A standardised terminological resource such as the BioLexicon (LexEBI) fulfils this need and provides additional information for the disambiguation of biomedical named entities.

The content from the scientific literature is processed undergoing entity, concept and fact extraction. The concept annotations for the entities are then integrated into the text according to the IeXML format:

<e id="Uniprot:P01308:T028:PRGE|UMLS:C1337112: T028: PRGE">INS gene</e>

The annotations of the entities make reference to the primary data resource of the entities and allow for ambiguous and nested annotations [30].

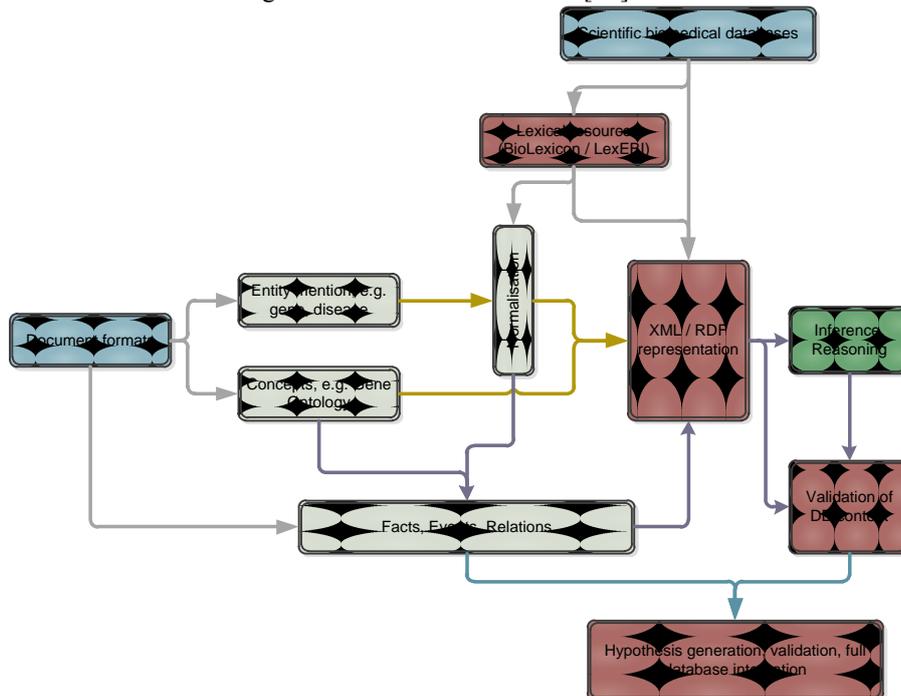

**Fig. 1.** Processing the scientific literature requires basic processing steps such as entry and concept recognition and leads into fact, event and relation extraction [32,33]. Entities are normalised to be linked to database concepts. A lexical resource contributes to this normalisation step. Thereafter, the content from the information extraction step is integrated into a triple store for further inference and reasoning, validation of the database content and hypothesis generation.

### 3.2 Querying the BioLexicon / LexEBI in the triple store

The terminological resource serves as normalising resource. The RDF representation is accessible in the triple store and links the terminological term variants to the primary data resources. Fig. 2 shows the example of the term variants of a UniProt entry. Statistical information is accessible to perform basic disambiguation. A single term variant from the scientific literature can be resolved through the lexical resource to one or several data entires in the biomedical data resource.

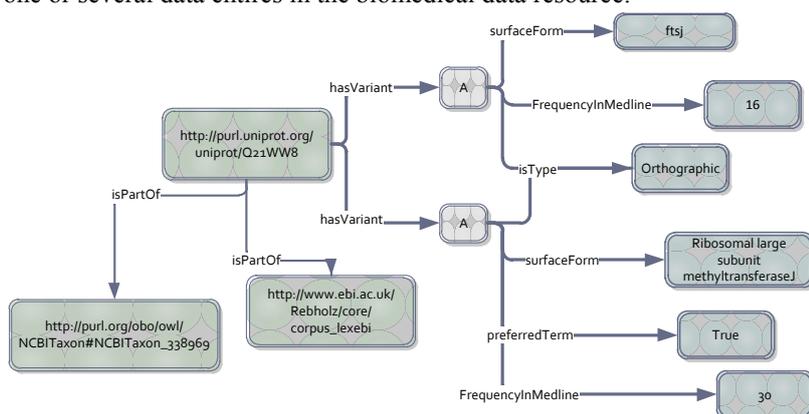

**Fig. 2.** The representation of the lexical item in the triple store gives access to the different surface forms of the term, i.e. the term variants, the frequency counts of the term, i.e. in this case on the frequency across Medline, and the reference to the primary conceptual resource for semantic interoperability.

### 3.3 Querying the CALBC content from the triple store

The annotations from the scientific literature are accessible as references from the document either only providing the term variant together with the semantic type, or as a concept reference to the lexical resource (see fig. 3). The latter case is required, if the retrieval uses only the concept identifier for interlinking of the literature content with the biomedical data resources. All meta-data from the scientific publication is accessible as well, such as the authors, the title and the journal of the publication and the data of publication. The meta-data information is represented following the Dublin Core initiative.

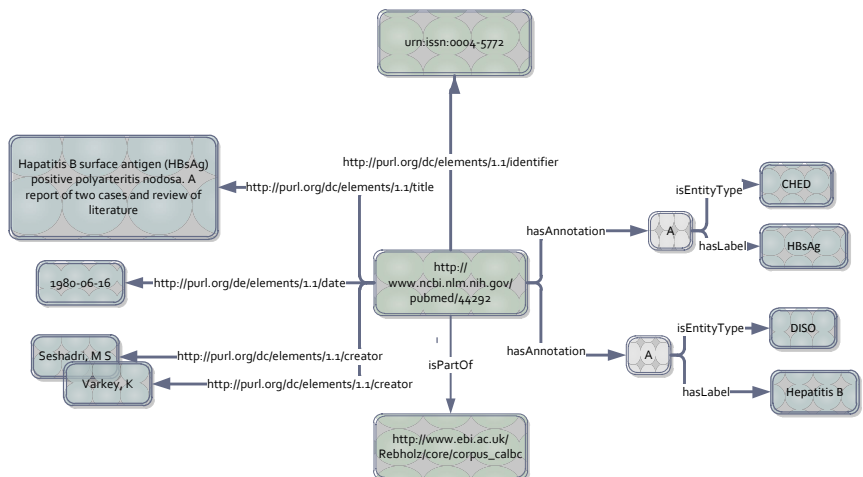

**Fig. 3.** The diagram shows the RDF representation of the annotations from a single annotated document. The Triple store enables access to the Meta-Data of the document, e.g. publication data, authors, and title of the manuscript, and on the other side access to the annotated content. In the current status only the labels of the entities are available from the text. The integration of the concept ids is ongoing work and requires improvements on the harmonisation of the concept annotation in the CALBC corpus.

In addition the annotations are referenced from the individual sentences indicating the position of the entities (see fig. 4). On the Sentence Level, the used URIs are local and specific to the CALBC project. Each sentence makes reference to at least one entity labelled with its semantic group (SPE, CHED, PRGE or DISO), the identified term in the corpus and the absolute position of the term in the sentence (including white spaces and tags).

This representation enables the identification of co-locations of entities in individual sentences and could be extended to more complex syntactical structures. For example, the BioLexicon / LexEBI and also other resources include references to verbs. The annotation of verbs and their nominalisations could be used for basic relation extraction on the sentence level.

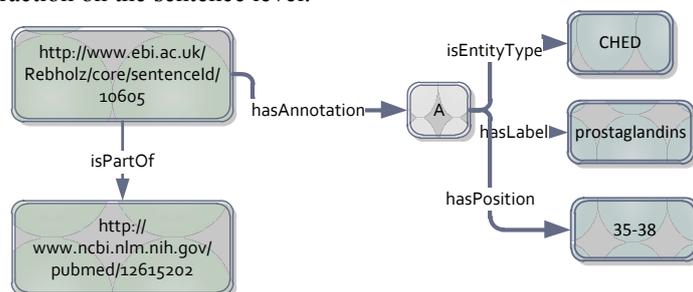

**Fig. 4.** The diagram shows the annotation on a sentence level similar to the document level shown in figure 1. Now the annotation is given together with the sentence position for accurate retrieval.

### 3.4 The UniProt triple store

The UniProt data resource (see fig. 5) provides links to other biomedical data resources that give additional information for the annotation of the protein entry. Some of the resources are integrated into Bio2RDF. The GO annotations and interaction information is relevant for our triple store.

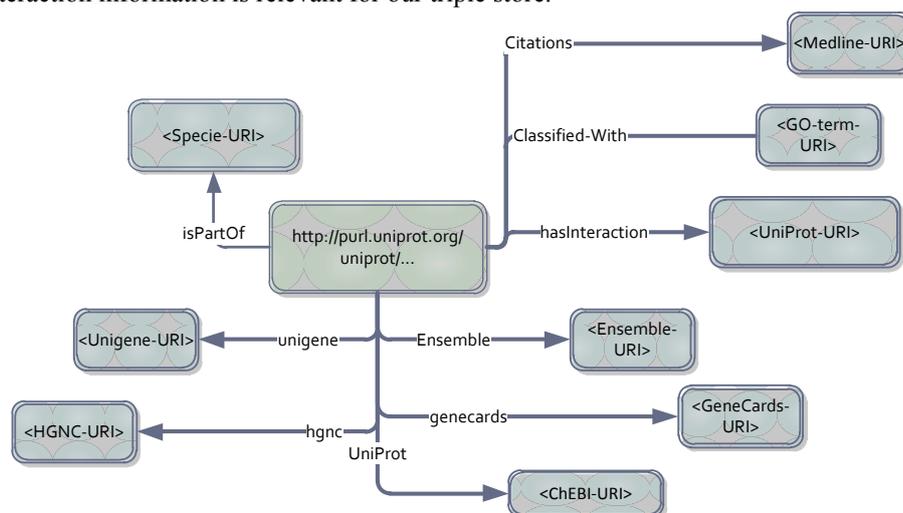

**Fig. 5.** The diagram gives an overview on the UniProt triple store that is used for the integration of the CALBC annotated corpus into the triple store. Several "same-as" relationships with different labels link the concept to related data resources such as Ensemble, GeneChards, ChEBI,, HGNC and Unigene. Not all mappings exist for all resources. Additional content from UniProt is providing GO classifications, interaction information, species annotations and references to Medline.

### 3.5 Querying the triple store

The LexEBI terminological resource makes reference to 1,178,659 clusters or unique concept ids, 3,848,775 terms, and 2,665,753 unique terms. The terminological resource can serve two different purposes: (1) mining the entities from the scientific literature and (2) improving the information retrieval from the annotated corpus.

The scientific literature has been integrated into the triple store from the CALBC corpus. The corpus represents a consensus annotation from different groups that have delivered annotations to the CALBC challenge. This approach enables the selection of annotations that are shared by different annotations solutions.

The SSC-II contains the following annotations: CHED 238,431, PRGE 435,797, DISO 245,524, and SPE 304,503. The content of the SSC-II has been fully integrated into the RDF Triple Store (4,568,678 triples). The UniProtKb triple store has been reduced to the content for human genes and proteins leading to about 12,552,239 triples for human. The integration of the content from GeneAtlas is ongoing work.

The UniProtKb triple store has been subselected for human genes only, makeing reference to 20,272 unique human gene entries in total. 7,598 distinct GO concepts are linked to the selected genes leading to 100,599 distinct GO concepts lead to 100,599 GO annotations. In addition, 13,897 interaction annotations have been extracted from the triple store. The generated ArrayExpress triple store contains references to 138 experiments. These experiments cover 15,135 distinct or unique genes.

RDF Triple Store enables querying the scientific literature and bioinformatics resources at the same time for evidence for gene-disease links that involve immunological processes. In total the CALBC RDF Triple Store makes use of 1,224,255 annotations in the corpus for exposing links between the entities supported by the evidence in the text. RDF Triple Store is implemented as a retrieval engine that allows querying for collocations of named entities and associated relevant information from the bioinformatics data resources (UniProtKb, ArrayExpress, see figure 6).

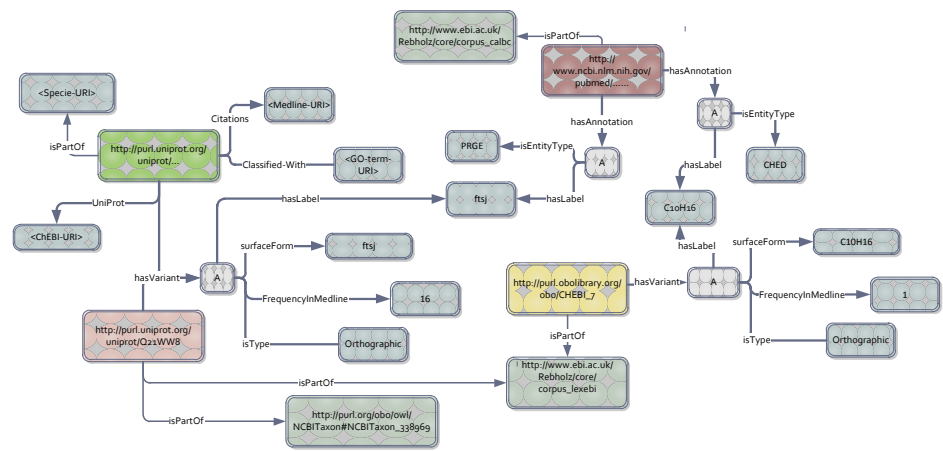

**Fig. 6.** The triple store brings together the scientific literature as Medline abstracts, entries from the lexical resource LexEBI and data from biomedical scientific databases. For clarity reasons, the links do not include the mention of the namespaces (lexebi, uniprot, calbc).

The integration of the lexical resource into the triple store enables to profit from lexical information when querying the triple store. For example the identified entity in the document can be disambiguated to a specific type based on the frequency information contained in the lexical resource. The following query gives another example of a query that uses the information contained in the lexicon to sort the retrieved documents according to the term frequency in the lexical resource.

Triple Store query:
```
PREFIX lexebi: <http://www.ebi.ac.uk/Rebholz/core/lexebi#>
PREFIX xsd: <http://www.w3.org/2001/XMLSchema#>
PREFIX owl: <http://www.w3.org/2002/07/owl#>
PREFIX rdf: <http://www.w3.org/1999/02/22-rdf-syntax-ns#>
PREFIX ebi: <http://www.ebi.ac.uk/Rebholz/core/>
PREFIX obo: <http://purl.obolibrary.org/obo/>
PREFIX expasy: <http://www.expasy.org/enzyme/>
```

```
PREFIX taxo: <http://purl.org/obo/owl/NCBITaxon#>
PREFIX interpro: <http://www.ebi.ac.uk/interpro/>
PREFIX umls: <http://url_umls#>
PREFIX uniprot: <http://purl.uniprot.org/uniprot/>
PREFIX pubmed: <http://www.ncbi.nml.nih.gov/pubmed/>
PREFIX dc: <http://purl.org/dc/elements/1.1/>
PREFIX calbc: <http://www.ebi.ac.uk/Rebholz/core/calbc#>

SELECT * WHERE {
  ?pmid calbc:hasAnnotation [calbc:hasLabel "String_to_query"] .
  ?lexebi_entity          lexebi:hasVariant         [lexebi:surfaceForm
"String_to_query", lexebi:frequencyInMedline ?mfreq].
}
ORDER BY DESC(?mfreq)
```

## 4  Discussion

In the first instance, the triple store supports retrieval that is based on co-occurrence. In addition and due to the implementation of the triple store, more complex retrieval functions are covered as well. First, the annotation of the entities with semantic types or concept ids resolves the ambiguity of the terms used in the text. Second, the information about the position of the entities in the sentence can be exploited to calculate a confidence score for the relatedness of entities in a given sentence.

The use of LexEBI as terminological resource provides additional benefits such as retrieval of evidence, where the preferred term for a named entity has been used or where the most or least frequent term variant for an entity is mentioned. Not all the data contained in LexEBI has been exploited. The cross-references between the different semantic types reach a complexity which cannot be fully met in the triple store. On the other side, some of the relations such as between UniProtKb and ChEBI are already covered by the other primary triple store content. Modelling LexEBI in OWL/RDF is an option to better support inference and full integration into semantic web solutions.

The CALBC and LexEBI triple stores have been implemented in Jena TDB following the results from a recent benchmark for performance . Jena TDB supports automated ontology reasoning (Racer, Pellet) that could be exploited across the Uniprot or ChEBI ontology.

## 5  Conclusion

Our triple store implementation brings together a large-scale annotated corpus, a lexical resource and public biomedical data resources. It is the first solution to retrieve content across all contained data including the scientific literature. The chosen approach enables basic identification of entity relations and forms the most common solution for the extraction of assertions and can be improved to identify more specific events. Overall the solution forms an open infrastructure for the validation of biomedical data resources against the literature, the validation of scientific hypotheses and the identification of hidden knowledge. Reasoning across all data resources would be the next large-scale improvement step.

**Acknowledgments.** This work was funded from the EU Support Action grant 231727 under the 7[th] EU Framework Programme within Theme "Intelligent Content and Semantics" (ICT 2007.4.2). The integration of the UniProt Triple Store and the ArrayExpress data was funded by the Pistoia Alliance as part of the SESL project.

# 5   The References Section